\documentclass{ws-procs9x6}

\usepackage{graphicx}
\DeclareGraphicsExtensions{.eps}
\usepackage{subfigure}

\newcommand{\note}[1]{\textbf{#1}}
\newcommand{\mathvec}[1]{\mbox{\boldmath$#1$}}

\begin{document}

\title{Towards a new measurement of the electron's electric dipole moment}

\author{J.~J. HUDSON, P.~C. CONDYLIS, H.~T. ASHWORTH, M.~R. TARBUTT, B.~E. SAUER, E.~A. HINDS}

\address{Centre for Cold Matter,\\
Imperial College London, SW7 2BW.\\
E-mail: ed.hinds@imperial.ac.uk}

\maketitle

\abstracts{\note{We present our progress towards a new measurement
of the electron electric dipole moment using a beam of YbF
molecules. Data are currently being taken with a sensitivity of
$10^{-27}\textrm{e.cm}/\sqrt{\textrm{day}}$. }}

\section{Motivation and historical perspective}

The most precise electric dipole moment (edm)
measurement\cite{Commins02} on the electron gives $d_e = (6.9 \pm
7.4)\times 10^{-28}\ \textrm{e.cm}$. Although the Standard Model
predicts a far smaller value, many modern extensions of particle
theory lead quite naturally to a value in the range of $10^{-27}\
\textrm{e.cm}$ or a little below\cite{CPVWS}. Our experiment,
which aims to be more sensitive than this, is therefore a search
for new physics. An edm at this level would also demonstrate a new
type of CP violation, beyond the usual CKM mechanism, as is
required to understand the matter-antimatter asymmetry of the
universe\cite{Sakharov}.

The interaction between $d_e$ and an applied field $\bf{E}$ can be
expressed by the effective non-relativistic Hamiltonian  $ -
d_e\,\alpha(E)\,\hat{\mathvec{\sigma}} \cdot \bf{E}$. For a free
electron, $\alpha(E) = 1$ and $\hat{\mathvec{\sigma}}$ is a unit
vector along the spin. If the electron is part of an atom or
molecule, $\hat{\mathvec{\sigma}}$ lies along the spin of the
system and $\alpha(E)$ is a factor that depends on the structure.
Some heavy atoms and molecules have the virtue that
$\alpha(E)\gg1$, and then it is called the enhancement
factor\cite{Sandars:atom}. This coupling resembles the interaction
$-\mu\,\beta(B)\,\mathvec{\hat{\sigma}} \cdot \bf{B}$ of the
magnetic moment $\mu$ with a magnetic field $\bf{B}$, where
$\beta(B)$ accounts for the atomic or molecular structure. It is
instructive to compare these two interactions in the case of a
free electron with an edm of, say, $d_e = 5\times10^{-28}\
\textrm{e.cm}$, just below the present limit. In a
$100\,\textrm{kV/cm}$ field the edm energy is so small that it
equals the magnetic energy in a field of only $9\times10^{-19}\
\textrm{T}$. Controlling the stray magnetic field at that level
seems close to impossible, especially when applying the electric
field. Heavy atoms such as Cs and Tl alleviate this problem by
their large enhancement factors. In particular, $\alpha(E) = -585$
for the thallium atom\cite{Liu}, which relaxes the necessary field
control to the challenging, but achievable fT level. Two magnetic
effects are most troublesome. (i) Stray magnetic fields vary both
in space and time. (ii) Atoms moving through the large electric
field experience a motional magnetic interaction\cite{AC paper}
$-\mu\,\mathvec{\hat{\sigma}}\cdot\bf{E\times v}/c^2$. In both
cases the unwanted field components are typically many orders of
magnitude larger than fT and heroic efforts were needed to reach
the current precision\cite{Commins02}.

\section{Measuring with molecules}
Heavy polar molecules offer substantial relief from these
difficulties\cite{physica scripta}. First, the enhancement factors
are generically much larger\cite{molecule enhancement} because the
electron edm interacts with the polarisation of the charge cloud
close to the heavy nucleus. In an atom this polarisation follows
from the mixing of higher electronic states by the applied
electric field. In a polar molecule, these electronic states are
already strongly mixed by the chemical bond and it is only
rotational states that have to be mixed by the applied field.
Since these are typically a thousand times closer in energy, the
molecular enhancement factor is larger. For the YbF molecule used
in our experiment, the enhancement factor\cite{Kozlov:1998} is
$\alpha \simeq 10^6$ at our operating field of $13\
\textrm{kV/cm}$, which relaxes the requirement on field control to
the $\textrm{pT}$ level.

There is a second advantage to YbF\footnote{or any system whose
tensor Stark splitting greatly exceeds the Zeeman interaction.}.
Being polar, this molecule has a strong tensor Stark splitting
between the Zeeman sublevels, making the applied electric field
the natural quantization axis. As a result, the Zeeman shift
associated with a perpendicular magnetic field is strongly
suppressed, making the molecule insensitive to the motional field.
For our typical operating parameters the motion-induced false edm
is\cite{Hudson} below $10^{-33}\,\textrm{e.cm}$, which is entirely
negligible.

\section{YbF spin interferometer}

\begin{figure}[htb]
\begin{center}
\includegraphics[width=8.5cm]{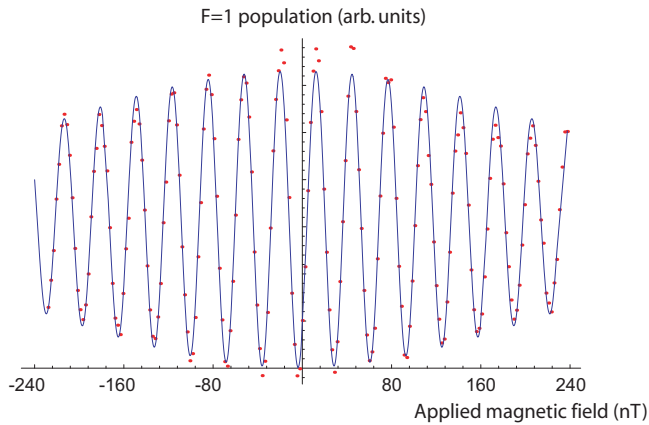}
\caption{YbF interferometer fringes. Dots: $F=1$ population
measured by fluorescence. Curve: Calculated fringes with signal
scale and offset in both coordinates as free parameters. Variation
of fringe visibility is due to the known beam velocity
distribution.} \label{fringes}
\end{center}
\end{figure}

Our experiment uses a cold, pulsed, supersonic beam of YbF
radicals\cite{Tarbutt} in a magnetically shielded vertical vacuum
chamber $\sim 1.5$m high. The electronic, rotational and
vibrational ground state $X^2\Sigma^+_{1/2},N=0,v=0$ is a
hyperfine doublet with states $F=0,1$ split by 170\,MHz. We first
deplete the $F=1$ state by laser excitation of the $F=1$ molecules
on the $A_{1/2}\leftarrow X$ transition. This laser beam is called
the pump. An oscillating field, which we call the first beam
splitter, then drives the $F=0$ molecules into a symmetric
coherent superposition of the $F=1, m_F = \pm 1$ states, as
described later in more detail. Next, parallel dc electric and
magnetic fields\footnote{The direction of these fields defines the
quantisation axis of our basis.} are applied to introduce a phase
shift $\Delta\phi = \frac{2}{\hbar}\int_0^\tau ( d_e\ \alpha(E)\
E(t) + \mu\ \beta(B)\ B(t)) \mathrm{d}t $ between the two
superposed states. Here $E$ and $B$ appear as functions of time
because they are the fields in the molecular rest
frame\footnote{Variation of field directions can generate
geometric phases, neglected here for simplicity.}. At time $\tau$
the molecules interact with another oscillating field, the
recombining beam-splitter, that couples the symmetric part of the
$F=1$ coherence back to the $F=0$ state. The resulting $F=0$ state
population exhibits the usual
$\cos^2\left(\frac{\Delta\phi}{2}\right)$ fringes of an
interferometer. We detect the $F=1$ population using fluorescence
induced by a probe laser on the $A_{1/2}\leftarrow X$ transition.
Figure\,\ref{fringes} shows the interference fringes observed in
this fluorescence when the magnetic field is scanned.

\section{The beam splitters}

\begin{figure}[tb]
\begin{center}
\includegraphics[width=7cm]{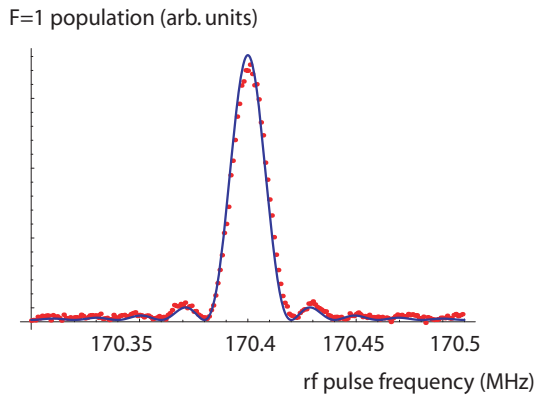}
\caption{Lineshape for rf beam splitter. Dots: experiment. Line:
eq.\ref{Rabi} with 50$\mu$s rf pulse length.} \label{rf_lineshape}
\end{center}
\end{figure}

We describe two types of beam splitter, both designed to minimise
the distance moved by the molecules whilst they are being split.
This is important as unwanted phase shifts can occur if the
ambient fields rotate during splitting. When combined with other
imperfections of the apparatus, such a phase can produce a false
edm and is therefore undesirable.

\subsection{Radio frequency splitter}
\begin{figure}[htb]
\centering
\subfigure[]{\includegraphics[width=5cm]{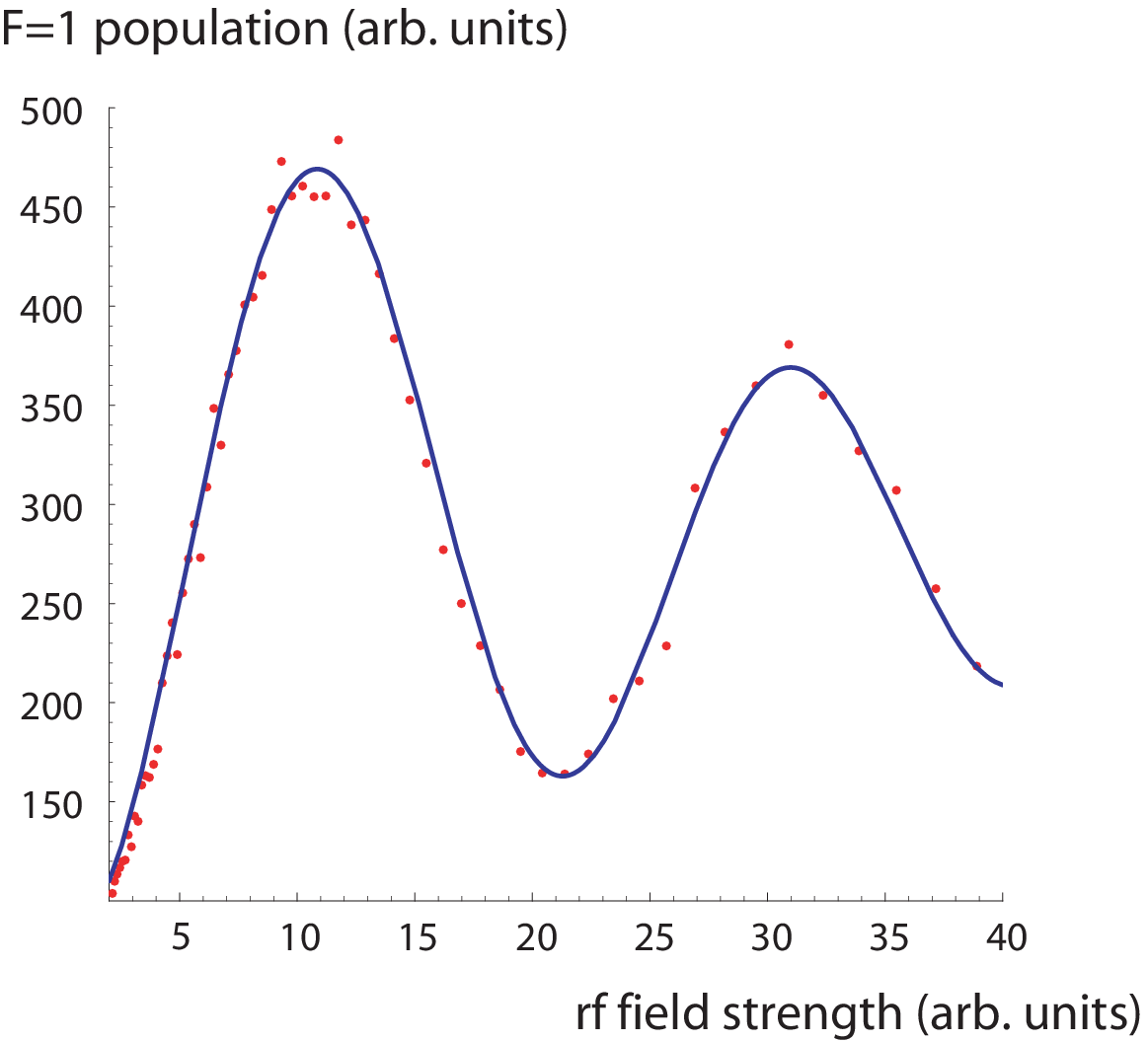}\label{damped_rabi}}\quad
\subfigure[]{\includegraphics[width=5cm]{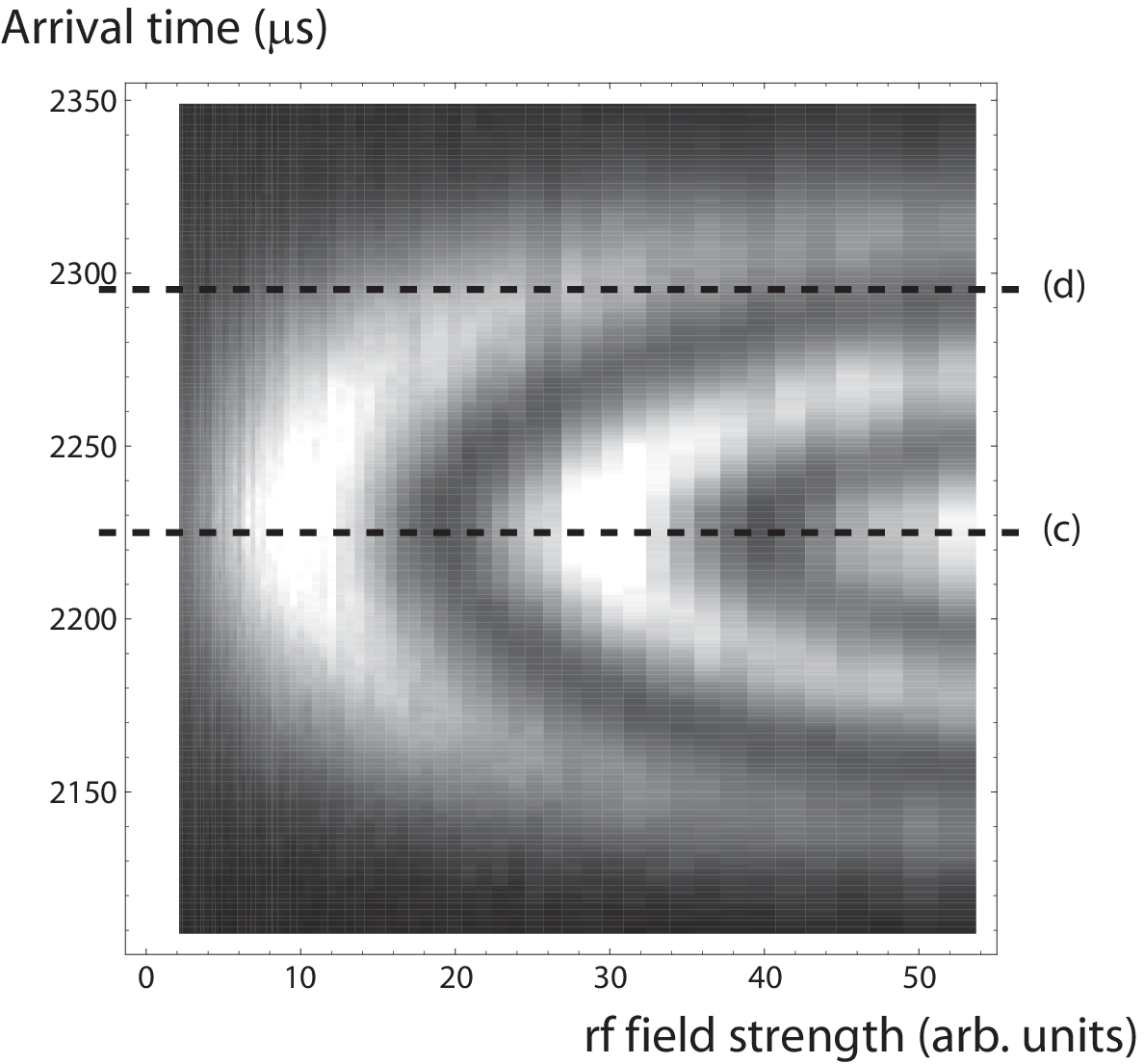}\label{rabi_density}}
\subfigure[]{\includegraphics[width=5cm]{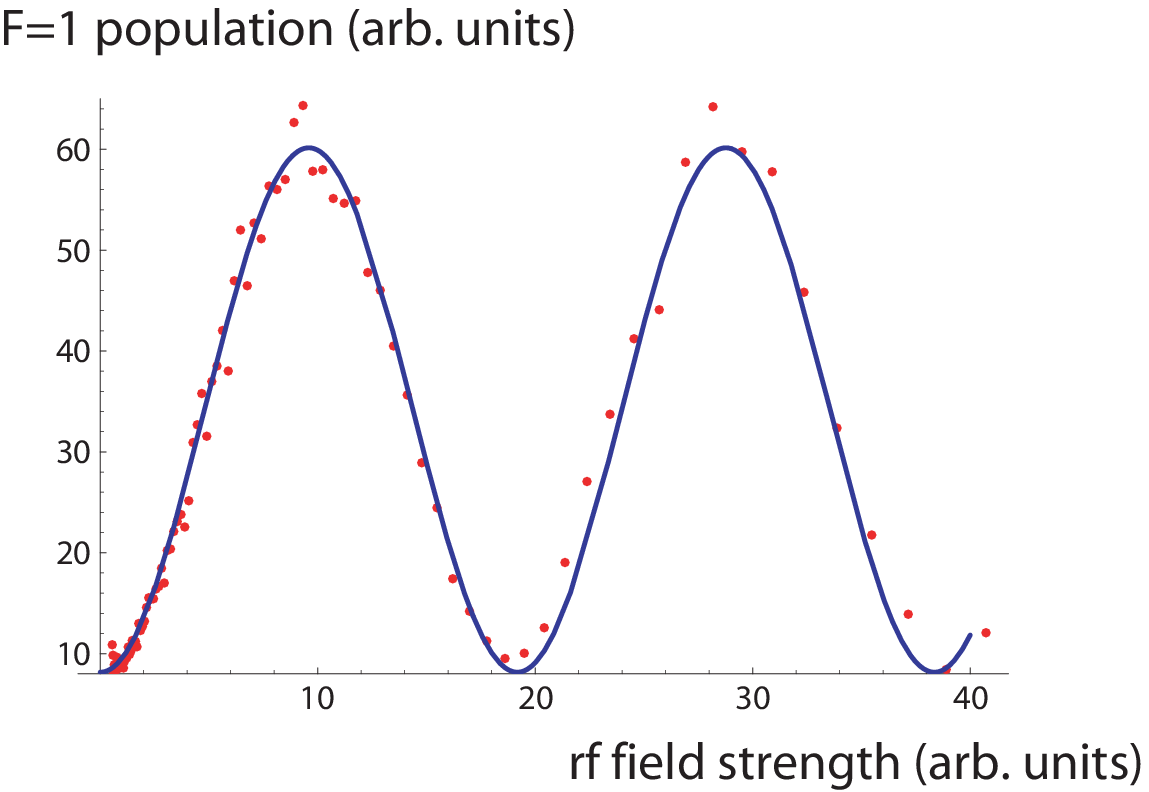}\label{rabi_slice1}}\quad
\subfigure[]{\includegraphics[width=5cm]{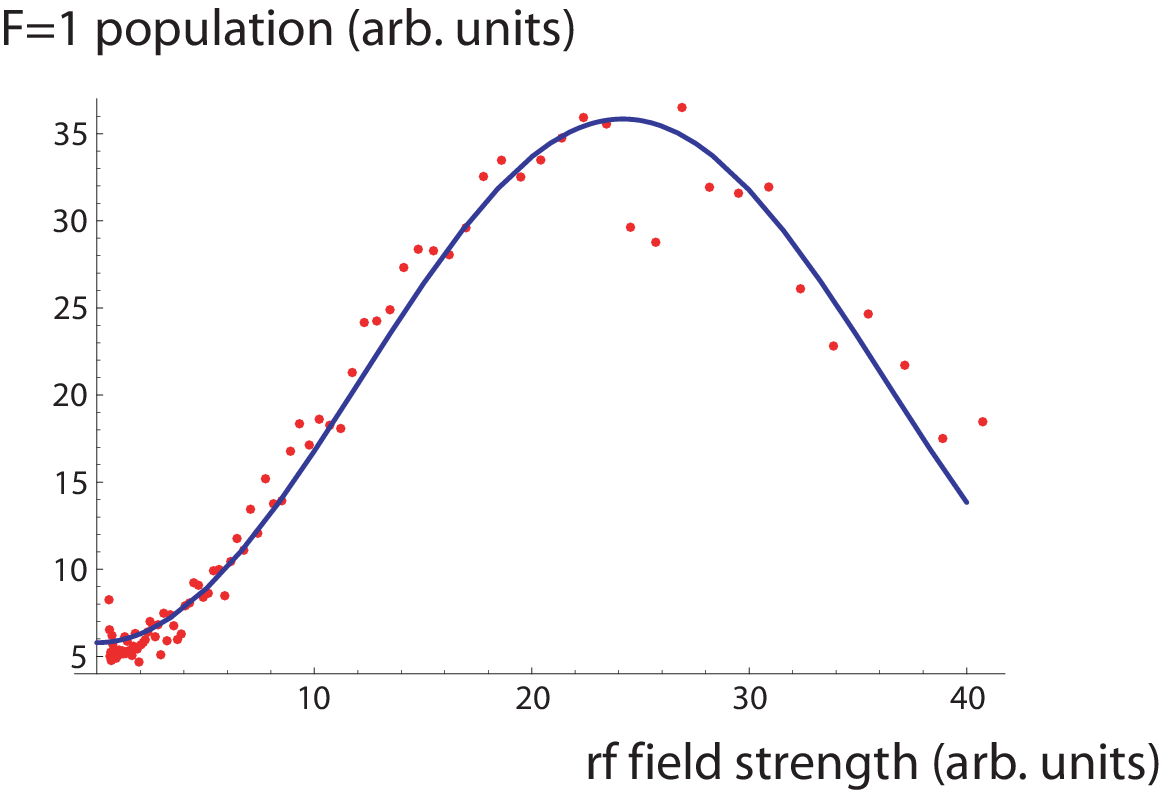}\label{rabi_slice2}}
\caption{Rabi flopping in the rf beam splitter. Units on axes of
all the plots are arbitrary but common. The $\pi$-pulse occurs at
a forward rf power to the loop of around 100mW. The dashed lines
in (b) show sections taken for (c) and (d).} \label{rabi}
\end{figure}

The radio frequency beam splitter is an rf magnetic field
perpendicular to $\bf{E}$ along the beam direction. This excites
the $F=1$ coherent superposition with probability\cite{Ramsey}
\begin{equation}
\label{Rabi} \textrm{P}_{\textrm{F}\rightarrow1} =
\frac{\Omega^2}{\delta^2 + \Omega^2} \sin^2\left( \frac{1}{2}
\sqrt{\Omega^2 + \delta^2}\ t\right)\ ,
\end{equation}
where $\Omega$ is the Rabi frequency and $\delta$ is the detuning
of the rf field from resonance. When a pulse of molecules arrives
at the centre of the rf loop we subject it to a short rf pulse
such that $\Omega t=\pi$, i.e. a $\pi$-pulse, to induce 100\%
population transfer at resonance. Figure\,\ref{rf_lineshape} shows
the lineshape measured using $50\,\mu$s-long rf pulses, together
with a fit to eq.\,\ref{Rabi}, showing quite good agreement.
However, the peak transition probability is significantly less
than unity. This is due to the beam velocity spread, which makes
the gas pulse 5\,cm long at the first loop and 10\,cm at the
second. Since the rf field strength varies along the beam line,
this spreading produces a distribution of Rabi frequencies. The
effect appears more clearly as a damping of the Rabi oscillations,
shown in fig.\,\ref{damped_rabi}, in good agreement with the
expected damping from the known magnetic field distribution of the
rf loop. The maximum fraction of the population transferred to
$F=1$ is 66\%\footnote{A longer rf $\pi$-pulse increases this
fraction by allowing each molecule to sample a longer distance
along the beamline, but this is avoided because of the possible
systematic errors due to large movements of the molecules.}.

The arrival of each YbF pulse at our detector is recorded with
$1\mu\textrm{s}$ resolution. This time-resolved data gives us a
spatial resolution of $\sim 5\ \textrm{mm}$ at the upper rf
loop\footnote{Note, however, that the molecules move approximately
$3\,\textrm{cm}$ during the 50\,$\mu$s rf pulse.}. This allows us
to resolve the inhomogeneous Rabi frequency, as shown in
fig.\,\ref{rabi_density}. The plot maps rf transition probability
(lighter shade meaning higher probability) versus rf field
strength and arrival time at the detector. Contours of constant
$\Omega t$ are clearly seen. At the centre of the loop,
corresponding to the dashed line (c) in fig.\,\ref{rabi_density},
we can drive more than two complete Rabi oscillations, as plotted
in fig.\,\ref{rabi_slice1}. Figure\,\ref{rabi_slice2} shows the
much slower Rabi flopping of late-arriving molecules (section (d)
of fig.\,\ref{rabi_density}) that experienced a weaker rf field.
The data follow the expected sinusoidal Rabi-flopping behaviour
(lines), demonstrating that the damping in fig.\,\ref{damped_rabi}
is indeed due to rf field inhomogeneity.

We are currently working to shorten the rf pulses further so that
the molecules move even less distance during the splitting.
Preliminary results with new, higher power, rf amplifiers suggest
that we can work with pulses less than $10\,\mu$s long,
corresponding to a beam movement below 6\,mm.

\subsection{Raman splitter}

The Raman beam splitter uses two co-propagating 552\,nm laser
beams with a frequency difference of $\sim170\,$MHz to drive the
hyperfine transition. In this case the spatial localisation of the
light ensures that the transition occurs in well-defined static
fields. An adequate hyperfine transition rate is achieved by
tuning the light near the $A_{1/2}\leftarrow X$ transition. At the
same time we avoid spontaneous emission from the $A_{1/2}$ state
as that tends to leave the molecule in a vibrationally or
rotationally excited state. We achieve the required detuning by
Stark shifting the $A_{1/2}\leftarrow X$ transition. This allows
us to use just one dye laser to generate the Raman beams as well
as the pump and probe light - a valuable simplification. We apply
an electric field of $2.7\ \textrm{kV/cm}$ to detune the
transition\cite{stark paper} by $\Delta=250\,\textrm{MHz}$, which
is $\sim10$ natural linewidths. The excitation probability is then
well approximated by eq.\,\ref{Rabi} with the substitution
$\Omega=\Omega_1 \Omega_2/\Delta$, where $\Omega_{1,2}$ are the
Rabi frequencies for the two laser fields. Now, $\delta$
(eq.\,\ref{Rabi}) is the detuning from two-photon resonance and
$t$ is the time of flight through the cw Raman splitter. Once
again, the velocity spread of the beam makes it impossible to give
all the molecules a $\pi$-pulse. However, the spread is only some
40$\,\textrm{m/s}$ FWHM on a mean velocity of $580\,\textrm{m/
s}$, allowing a transition probability of 99.7\% to be achieved.

\begin{figure}[htb]
\centering
\subfigure[]{\includegraphics[width=5cm]{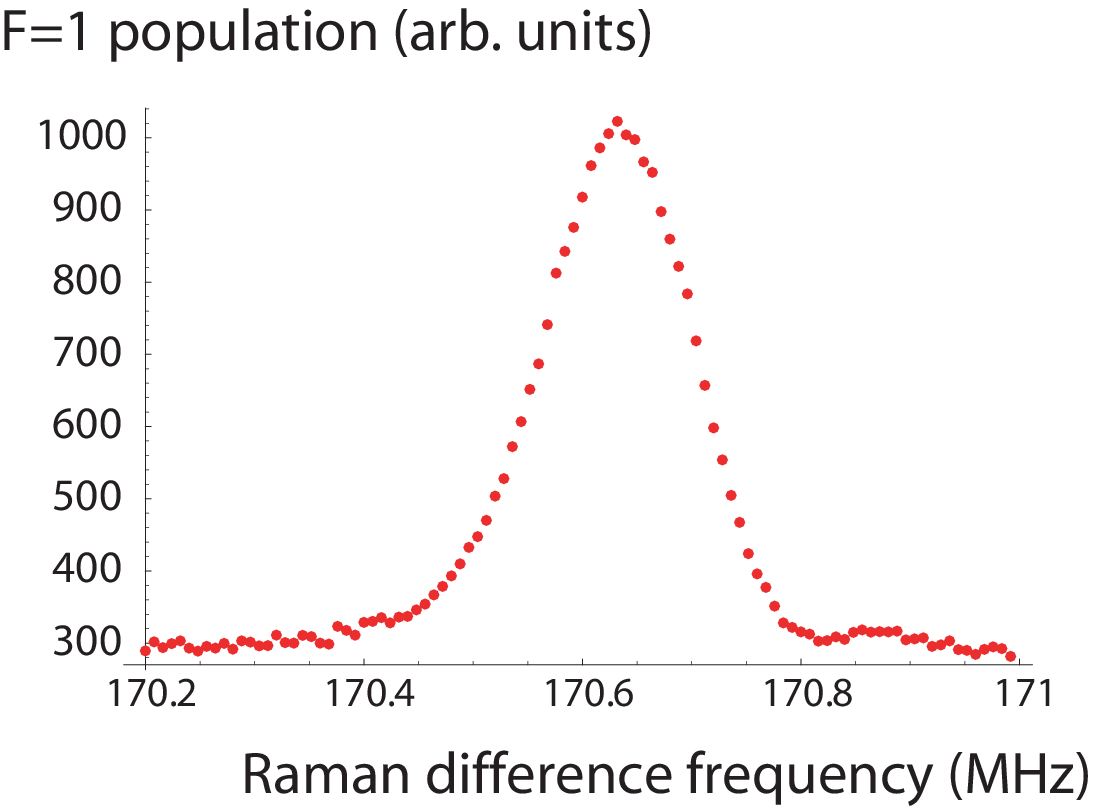}\label{raman_lineshape}}\quad
\subfigure[]{\includegraphics[width=5cm]{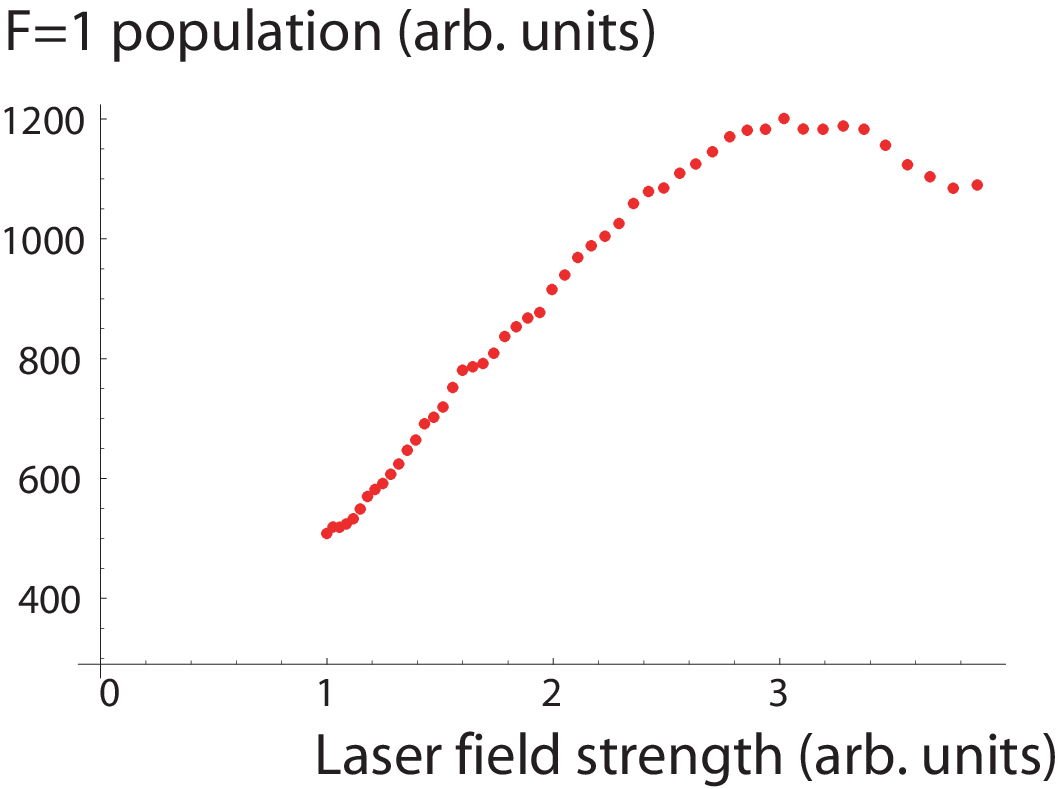}\label{raman_flopping}}
\caption{Raman beam splitter. (a) Transition lineshape. (b) Rabi
flopping on the Raman beam splitter transition.} \label{raman_fig}
\end{figure}

Figure\,\ref{raman_lineshape} shows the Raman transition
lineshape. The width of this line is dominated by Doppler
broadening from the transverse momentum spread of the laser beams,
an effect that is less well controlled than we would like in our
edm measurement. In fig.\,\ref{raman_flopping} we show the Rabi
flopping of the transition, with the intensity held fixed in one
beam (the redder one) and varied in the other. Limited laser power
restricts us to less than a full cycle.

\section{Results and outlook}

\begin{figure}[htb]
\begin{center}
\includegraphics[width=12cm]{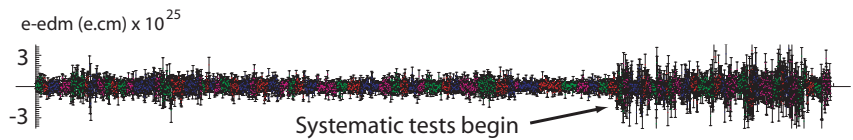}
\caption{The current data set. Each of the 3088 points is an
independent electron edm measurement over approximately two
minutes. 
} \label{data}
\end{center}
\end{figure}

We are now making an electron edm measurement. Our current data
set, collected over the last two months is displayed in
chronological order in fig.\,\ref{data}. There are 3088
measurements each of which took typically 2 minutes. The small
variations in average sensitivity are due primarily to changes in
the intensity and stability of the molecular beam and, to a lesser
degree, of the laser. In the last two weeks there is a period of
sharply increased uncertainty as we start to make checks for
systematic errors.

Although we cannot present a value for the edm at this conference,
we can make some comments on our data set. Since our last edm
measurement\cite{Hudson}, gains from a cold, pulsed
source\cite{Tarbutt} and other technical improvements have
increased the experimental sensitivity by more than a factor of
thirty so that the statistical sensitivity is approximately
$10^{-27}\textrm{e.cm}/\sqrt{\textrm{day}}$. All the data were
taken using Raman beam splitters. We are preparing to take new
data with the pulsed rf beam splitters once the current systematic
checks are complete. As a further control against any false edm we
plan to repeat the experiment using CaF molecules. These provide a
good null test as CaF is similar to YbF structurally and
magnetically, but has $\sim 40$ times less sensitivity to the
electron edm according to the expected $Z^3$ scaling\cite{physica
scripta}. We have already performed a trial run and the results
look encouraging for making a measurement in the
$10^{-28}\textrm{e.cm}$ range in the near future.  Beyond that, we
plan to guide and decelerate the beam\cite{decelerator}, to obtain
an anticipated further factor of 100 in sensitivity.

We acknowledge support from PPARC, EPSRC, the Royal Society and
the Cold Molecules Research Training Network of the European
Commission. We are indebted to the European Science Foundation for
their conference series and to Jon Dyne for expert technical
assistance.

\end{document}